\documentclass[11pt]{iopart}

\usepackage[utf8]{inputenc}

\usepackage[T1]{fontenc}

\usepackage{epsfig}
\usepackage{bm}
\usepackage{color}
\usepackage{verbatim}

\usepackage{iopams}
\expandafter\let\csname equation*\endcsname\relax
\expandafter\let\csname endequation*\endcsname\relax
\usepackage{amsmath}
\usepackage{amssymb}

\usepackage{bbm}
\usepackage{cite}

\newcommand{\bfx}{\mathbf{x}}
\newcommand{\bfX}{\mathbf{X}}
\newcommand{\bfC}{\mathbf{C}}
\newcommand{\bfI}{\mathbf{I}}
\newcommand{\bfU}{\mathbf{U}}
\newcommand{\bfe}{\mathbf{e}}
\newcommand{\bfSig}{\mathbf{\Sigma}}

\let \Im \relax
\DeclareMathOperator{\Im}{Im}

\begin{document}

\title[Nonequilibrium quantum fluctuation relations]{Nonequilibrium quantum
fluctuation relations for harmonic systems in nonthermal environments}

\author{D Pagel$^{1}$, P Nalbach$^{2}$, A Alvermann$^{1}$, H Fehske$^{1}$ and M
Thorwart$^{2}$}

\address{$^1$Institut f\"{u}r Physik, Ernst-Moritz-Arndt-Universit\"{a}t 
Greifswald, Felix-Hausdorff-Str.\ 6, 17489 Greifswald, Germany}
\address{$^2$I. Institut f\"ur Theoretische Physik, Universit\"at Hamburg,
Jungiusstra{\ss}e 9, 20355 Hamburg, Germany}

\ead{pagel@physik.uni-greifswald.de}

\begin{abstract}
We formulate exact generalized nonequilibrium fluctuation relations for
the quantum mechanical harmonic oscillator coupled to multiple harmonic baths.
Each of the different baths is prepared in its own individual (in general
nonthermal) state.
Starting from the exact solution for the oscillator dynamics
we study fluctuations of the oscillator position as well as of the energy
current through the oscillator under general nonequilibrium conditions.
In particular, we formulate a
fluctuation-dissipation relation for the oscillator position
autocorrelation function
that generalizes the standard result for the case of a single bath at thermal
equilibrium. 
Moreover, we show that the generating function for the position operator 
fullfills a generalized Gallavotti-Cohen-like relation. For the energy
transfer through the oscillator, we determine the average energy current
together with the current fluctuations.
Finally, we discuss the generalization of the cumulant generating function for
the energy transfer to nonthermal bath preparations.
\end{abstract}



\maketitle

\section{Introduction}
Fluctuation relations 
\cite{Campisi11,Jarzynski11,Esposito09,Jarzynski08,Marconi08,Seifert08,
Rondoni07} build on the fundamental connection between the response of a
physical system to a weak externally applied force and the fluctuations in the
system
without the external force. This connection was first observed for thermal
equilibrium
by William Sutherland \cite{Sutherland02,Sutherland05} and
Albert Einstein
\cite{Einstein05,Einstein06a,Einstein06b}. They established the relation 
between the mobility of a Brownian particle, which is a quantity that measures
the response to an external electric field, and the diffusion constant, which is
a quantity that characterizes the fluctuating forces at equilibrium.
The famous Johnson-Nyquist relation
\cite{Johnson28,Nyquist28} gives the corresponding connection between
the electrical resistance of a circuit and charge fluctuations in the resistor.
A more general relation has been derived by Callen and
Welton \cite{Callen51} in form of the quantum fluctuation-dissipation theorem
(FDT) 
\begin{equation}\label{thermal0}
 \Psi(\omega) =  \frac{\hbar}{2 \rmi} \coth\Big(\frac{\hbar \beta
\omega}{2}\Big)
 \Phi(\omega) \:,
\end{equation}
which relates the Fourier transform $\Psi(\omega)$ of the symmetric
equilibrium correlation function of an observable to the Fourier transform 
$\Phi(\omega)$ of the (antisymmetric) response function of this observable in
thermal equilibrium at temperature $T=(k_B\beta)^{-1}$. It was
recognized by Green \cite{Green52,Green54} and Kubo \cite{Kubo1957} that the
FDT in Eq.\ (\ref{thermal0}) is a particular case of the more general linear
response theory which is an invaluable tool to model and understand 
experimental data in all fields of physics. However, often situations are 
encountered where the assumption of thermal equilibrium is invalid, for example,
for systems strongly driven by external fields, charge currents in systems with
large differences in the electric potential, heat currents in systems with
strong temperature gradients, or systems in solvents and disordered media which
themselves are in metastable quasi-equilibria only. It has been a longstanding
task in
statistical physics to generalize linear response theory 
and FDTs to such nonequilibrium situations and, by this, to build a unifying
theoretical framework of the spectral characteristics of environmental noise.

Generalized nonequilibrium fluctuation theorems have been formulated for
classical
nonstationary Markov processes \cite{Hanggi75} and 
for stationary Markov processes far away from thermal equilibrium
\cite{Hanggi78,Hanggi82}. They relate the higher-order nonlinear response to
higher-order correlation functions of stationary nonequilibrium fluctuations.  
A fully nonlinear, exact and universal classical fluctuation relation has been
provided by Bochkov and Kuzovlev \cite{Bochkov77}. It gives the fluctuation
relation at any order for systems that are in a thermal state in absence of
external forces. It solely builds on the time-reversal invariance of the
equations of motion and the assumption of a thermally equilibrated initial
state. The quantum version was provided by Andrieux and Gaspard
\cite{Andrieux08} and lead to fundamental insights \cite{Campisi11} into
the fact that work injected to or extracted from a system is
not a quantum mechanical operator or observable, because it characterizes a
process rather than a state of the system \cite{Talkner07}. 

Recently, growing interest in nonequilibrium fluctuation relations arose
from alternative formulations by Evans {\it et al.} \cite{Evans93}
and by Gallavotti and Cohen \cite{Gallavotti95} for the
statistics of nonequilibrium fluctuations in steady states and by 
Jarzynski \cite{Jarzynski97} and Crooks \cite{Crooks99} on the statistics of
work performed by a transient time-dependent perturbation
\cite{Campisi11}. The reviews 
\cite{Campisi11,Jarzynski11,Esposito09,Jarzynski08,Marconi08,Seifert08,
Rondoni07} summarize the actual progress in this field.

Most studies so far consider systems initially in
thermal equilibrium, described by the canonical
distribution 
\begin{equation}
 \rho_0=\frac{1}{Z_0} \rme^{-\beta H_0} \:,
\end{equation}
with the system Hamiltonian $H_0$ and the partition function $Z_0=\mathop{{\rm
Tr}}
[\rme^{-\beta H_0}]$. In this work, we want to give up this assumption and 
formulate generalized nonequilibrium fluctuation relations for nonthermal
initial states.
To do so, we consider the dissipative quantum mechanical harmonic oscillator 
\cite{Ullersma66,Haake85,Ford85,Ford88,Grabert88,Zerbe96,Thorwart00,Grabert06,
Ford07}.
Building on our previous work in Ref.\ \cite{Pagel2013} 
we study a central oscillator coupled to an arbitrary number of harmonic baths
each of which can be prepared in its own individual initial state.
The fluctuations of the baths are thus still Gaussian, but not necessarily
thermally distributed.
Because the exact solution for the system dynamics is known, 
we can analytically calculate all observables and correlation functions of
interest,
and thus investigate the validity of nonthermal nonequilibrium fluctuation
relations for this admittedly restricted model situation.

The structure of the paper is as follows. We introduce the model, its 
classical equation of motion and the basic notions in
Sec.~\ref{sec:Model}. Then, in Sec.\ \ref{sec:oscpos}, we calculate the
symmetric and antisymmetric correlation functions of the oscillator position
for the case of general nonthermal bath states. In Sec.\ \ref{sec:nfrpos}, we
formulate the generalized nonequilibrium fluctuation relation for the
oscillator position correlation functions. This constitutes one major result of
this work. In Sec.\ \ref{sec:genfuncpos}, we calculate the generating function
for the position operator of the oscillator and show that it fullfills a
generalized Gallavotti-Cohen relation under nonequilibrium conditions at
arbitrary times. Sec.\ \ref{sec.heat} is devoted to energy transfer and
 we present the derivation of the average energy current. In Sec.\
\ref{sec.heatflucs}, we calculate the energy current fluctuations and generalize
the well-known cumulant generating function of the heat transfer for thermal
baths to general bath preparations, before we summarize in
Sec.\ \ref{sec:concl}.

\section{The model} \label{sec:Model}
In a system-bath model approach, we consider the one-dimensional
 harmonic oscillator bilinearly coupled to a finite number $N_B$ of different
and mutually uncoupled baths of harmonic oscillators.
The total Hamiltonian is $H = H_S + H_B + H_{SB}$, where ($\hbar=1, k_B=1$
throughout the work)
\begin{equation}
 H_S = \frac{1}{2} \Big[ P^2 + \Omega^2 Q^2 \Big]
\end{equation}
is the contribution of the central oscillator with frequency $\Omega$,
\begin{equation}
 H_B = \sum_{\alpha=1}^{N_B} H_B^\alpha \;, \qquad
 H_B^\alpha = \sum_{\nu=1}^{N_\alpha} \frac{1}{2}
 \Big[ (P_\nu^\alpha)^2 + (\omega_\nu^\alpha Q_\nu^\alpha)^2 \Big]
\end{equation}
is the contribution of the bath oscillators with frequencies
$\omega_\nu^\alpha$, and
\begin{equation}
 H_{SB} = \sum_{\alpha=1}^{N_B} H_{SB}^\alpha \;, \qquad
 H_{SB}^\alpha = Q \sum_{\nu=1}^{N_\alpha} \lambda_\nu^\alpha Q_\nu^\alpha
 + \sum_{\nu=1}^{N_\alpha} \frac{1}{2} \Big( \frac{\lambda_\nu^\alpha}
 {\omega_\nu^\alpha} \Big)^2 Q^2
\end{equation}
is the coupling part.
In these expressions, the position and momentum operators $Q_\nu^\alpha$ and 
$P_\nu^\alpha$ fullfill the canonical commutation relation $[Q_\nu^\alpha,
P_\mu^{\alpha'}] = \rmi \delta_{\nu\mu} \delta_{\alpha\alpha'}$.
The labels $\alpha, \alpha' = 1, \ldots, N_B$ are used to identify a particular 
bath, while the indices $\nu, \mu = 1, \ldots, N_\alpha$ identify a single
oscillator from bath $\alpha$.

The coupling term contains the counter term which serves
to eliminate the potential renormalization due to the coupling of the 
oscillator to the baths \cite{Weiss2008,counterHanggi1997}. 
Throughout this work, we assume factorizing initial states $\rho(0) = \rho_S(0)
 \bigotimes\limits_{\alpha=1}^{N_B} \rho_B^\alpha(0)$ corresponding to the
choice of isolated systems that are brought into contact at $t = 0^+$.
Notice, however, that we keep the initial distributions $\rho_B^\alpha(0)$ of 
the baths arbitrary and do not necessarily assume thermal equilibrium.

\subsection{The exact solution for the operator dynamics}
Starting from the Heisenberg equation of motion for the system and bath
operators, one inserts the formal solution for the bath operator dynamics into
the equation of motion of the central oscillator to obtain the quantum Langevin
equation
\begin{equation}\label{QLE}
 \ddot{Q}(t) = -\Omega^2 Q(t) - \int_0^t \rmd\tau \: K(t - \tau)
\dot{Q}(\tau) - \eta(t) - K(t) Q(0)
\end{equation}
with the damping or friction kernel
\begin{equation}\label{K}
 K(t) = \sum_{\alpha=1}^{N_B} K_\alpha(t) \;, \qquad
 K_\alpha(t) = \sum_{\nu=1}^{N_\alpha} \Big(
\frac{\lambda_\nu^\alpha}{\omega_\nu^\alpha} \Big)^2 \cos \omega_\nu^\alpha t
\;,
\end{equation}
and the noise term
\begin{equation}
 \eta(t) = \sum_{\alpha=1}^{N_B} \eta_\alpha(t) \;, \qquad
 \eta_\alpha(t) = \sum_{\nu=1}^{N_\alpha} \lambda_\nu^\alpha \Big(
Q_\nu^\alpha(0) \cos \omega_\nu^\alpha t  + P_\nu^\alpha(0) \frac{\sin
\omega_\nu^\alpha t}{\omega_\nu^\alpha} \Big) \;.
\end{equation}
The noise term $\eta(t)$ together with the initial
slip term $K(t) Q(0)$ appears as a fluctuating force in Eq.~\eqref{QLE}. 
Due to our choice of factorizing initial states, the noise terms of different
baths are uncorrelated, i.\,e., $\langle \eta_\alpha(t) \eta_\beta(s) \rangle =
\langle \eta_\alpha(t) \rangle \langle \eta_\beta(s) \rangle$ for $\alpha \neq
\beta$. Nevertheless, the fluctuating forces $\xi_\alpha(t) = \eta_\alpha(t) +
K_\alpha(t) Q(0)$ including the initial slip term are correlated because of the
coupling to the central oscillator~\cite{Hanggi2005}. 
These correlations vanish if the expectation values are calculated with respect to the non-factorizing initial state that is obtained out of $\rho(0)$ through the unitary transformation with the displacement operator $\exp[\rmi Q \sum_{\nu=1}^{N_\alpha} \lambda_\nu^\alpha P_\nu^\alpha / (\omega_\nu^\alpha)^2]$.
At this point, we note that explicit expressions for the correlation
functions of the fluctuating forces depend on the choice of the initial
distributions $\rho_B^\alpha(0)$ and thus, the fluctuations are in general
associated with a nonstationary Gaussian operator noise. Only in the
limit of long times, these fluctuations become stationary again (see
\ref{CF_flukForce}).

As is well established in the literature~\cite{Ullersma66,Haake85,Ford85,Ford88,
Grabert88,Zerbe96,Thorwart00,Grabert06,Ford07}, the full solution for the
central oscillator dynamics can be
constructed from the solution $u(t) \in \mathbb{R}$ of the corresponding
classical equation of motion,
\begin{equation}
 \ddot{u}(t) = -\Omega^2 u(t) - \int_0^t \rmd\tau \: K(t-\tau) \dot{u}(\tau) \;.
\end{equation}
The relevant solution $u(t)$ is specified by $u(t) = 0$ for $t < 0$ and by the
initial conditions $u(0) = 0$ and $\dot{u}(0) = 0$.
It is given by the Fourier transform
\begin{equation}\label{uFromFourier}
 u(t) = \frac{1}{2 \pi} \int_{-\infty}^\infty \rmd\omega \: \rme^{-\rmi (\omega
+ \rmi 0^+) t} F(\omega + \rmi 0^+)
 = \frac{2}{\pi} \int_0^\infty \rmd\omega \: \sin \omega t \, \Im F(\omega +
\rmi 0^+)
\end{equation}
of the function
\begin{equation} \label{F}
 F(z) = \bigg[ \Omega^2 + \sum_{\alpha=1}^{N_B} \sum_{\nu=1}^{N_\alpha} \Big(
\frac{\lambda_\nu^\alpha}{\omega_\nu^\alpha} \Big)^2 - z^2 +
\sum_{\alpha=1}^{N_B} \sum_{\nu=1}^{N_\alpha} \frac{(\lambda_\nu^\alpha)^2}{z^2
- (\omega_\nu^\alpha)^2} \bigg]^{-1} \;.
\end{equation}

Given $u(t)$, the solution for the dynamics of the central oscillator operators
can be obtained from the matrix equation
\begin{equation}\label{qsolution}
 \begin{pmatrix} Q(t) \\ P(t) \end{pmatrix} = \bfU(t) \begin{pmatrix} Q(0) \\
 P(0) \end{pmatrix} - \sum_{\alpha=1}^{N_B} \sum_{\nu=1}^{N_\alpha}
\lambda_\nu^\alpha \bfU(t, \omega_\nu^\alpha) \begin{pmatrix} Q_\nu^\alpha(0) \\
P_\nu^\alpha(0) \end{pmatrix} \:.
\end{equation}
We here introduced the matrices
\begin{equation} \label{U}
 \bfU(t) = \begin{pmatrix} \dot{u}(t) & u(t) \\
 \ddot{u}(t) & \dot{u}(t) \end{pmatrix} \;,
\end{equation}
\begin{equation}
 \bfU(t, \omega) = \begin{pmatrix} u_R(t,\omega) & \dfrac{u_I(t,\omega)}{\omega}
 \\[2ex] v_R(t, \omega) & \dfrac{v_I(t, \omega)}{\omega} \end{pmatrix} \;,
\end{equation}
and denote by the respective index $R$ or $I$ the real or imaginary part of the
partial Fourier transforms of the classical solution $u(t)$,
\begin{equation}\label{partfourier}
 {u}(t, \omega) \equiv {u}_R(t, \omega) + \rmi{u}_I(t, \omega)
 = \rme^{\rmi \omega t} \int_0^t \rmd\tau \: u(\tau)
 \, \rme^{-\rmi \omega \tau} \;,
\end{equation}
\begin{equation}
 {v}(t, \omega) \equiv {v}_R(t, \omega) + \rmi{v}_I(t, \omega)
 = \rme^{\rmi \omega t} \int_0^t \rmd\tau \: \dot{u}(\tau)
 \, \rme^{-\rmi \omega \tau} = u(t) + \rmi \omega {u}(t, \omega) \;.
\end{equation}

\subsection{Expectation values}
Equation~\eqref{qsolution} allows us to express central oscillator  expectation
values for $t \geq 0$ in terms of the initial ones at $t = 0$.
The linear expectation values are given by the equation
\begin{equation}
 \bfX(t) \equiv \begin{pmatrix} \langle Q(t) \rangle \\ \langle P(t) \rangle
\end{pmatrix} = \bfU(t) \bfX(0) + \bfI(t) \;,
\end{equation}
where
\begin{equation} \label{I}
 \bfI(t) = -\sum_{\alpha=1}^{N_B} \sum_{\nu=1}^{N_\alpha} \lambda_\nu^\alpha
\bfU(t, \omega_\nu^\alpha) \bfX_\nu^\alpha
\end{equation}
depends on the initial bath expectation values $\bfX_\nu^\alpha = \big( \langle
Q_\nu^\alpha(0) \rangle, \langle P_\nu^\alpha(0) \rangle \big)^T$.

For the quadratic expectation values we define the correlator of 
two operators $A$ and $B$ by 
\begin{equation}
 \Sigma_{AB} = \frac{1}{2}\langle AB + BA \rangle - \langle A \rangle \langle B
\rangle \;,
\end{equation}
and write $\Sigma_{AB}(t) \equiv \Sigma_{A(t)B(t)}$ for better readability.
For correlators of operators related to bath oscillators at initial time, we
define
\begin{equation}
 \sigma_{A_\nu^\alpha B_\mu^\alpha} = \Sigma_{A_\nu^\alpha B_\mu^\alpha}(0)
\end{equation}
and write
\begin{equation}
 \bfSig_{\nu\mu}^\alpha = \begin{pmatrix} \sigma_{Q_\nu^\alpha Q_\mu^\alpha} &
\sigma_{Q_\nu^\alpha P_\mu^\alpha} \\ \sigma_{P_\nu^\alpha Q_\mu^\alpha} &
\sigma_{P_\nu^\alpha P_\mu^\alpha} \end{pmatrix} \;.
\end{equation}
We then obtain with Eq.~\eqref{qsolution} the relation
\begin{equation}
 \bfSig(t) \equiv \begin{pmatrix} \Sigma_{QQ}(t) & \Sigma_{QP}(t) \\
 \Sigma_{QP}(t) & \Sigma_{PP}(t) \end{pmatrix} = \bfU(t) \bfSig(0) \bfU^T(t) +
\bfC(t) \;,
\end{equation}
where
\begin{equation} \label{C}
 \bfC(t) = \sum_{\alpha=1}^{N_B} \sum_{\mu,\nu=1}^{N_\alpha} \lambda_\nu^\alpha
 \lambda_\mu^\alpha \bfU(t, \omega_\nu^\alpha) \bfSig_{\nu\mu}^\alpha
 \bfU^T(t, \omega_\mu^\alpha) \;.
\end{equation}

\subsection{The thermodynamic limit}\label{sec:tdLimit}
In the thermodynamic limit $N_\alpha \to \infty$ for all $\alpha = 1, \ldots,
 N_B$ we can replace summations $(1 / N_\alpha) \sum_{\nu=1}^{N_\alpha}
f(\omega_\nu^\alpha)$ by integrations $\int_0^\infty \rmd\omega \:
D_\alpha(\omega) f(\omega)$ by introducing the densities of states of the baths,
\begin{equation}
 D_\alpha(\omega) = \frac{1}{N_\alpha} \sum_{\nu=1}^{N_\alpha} \delta(\omega -
\omega_\nu^\alpha) \;,
\end{equation}
that converge to continuous functions.
Since the coupling constants $\lambda_\nu^\alpha$ enter Eqs.~\eqref{K} and 
\eqref{F} as $(\lambda_\nu^\alpha)^2$, they have to scale as $1 /
\sqrt{N_\alpha}$ to obtain finite results for the sum over $N_\alpha$ terms.
We thus introduce continuous functions $\lambda_\alpha(\omega)$ according to 
\begin{equation}
 \lambda^\alpha_\nu = \lambda_\alpha(\omega_\nu^\alpha) / \sqrt{N_\alpha} \;, 
\end{equation}
and define the bath spectral functions
\begin{equation} \label{specfunc}
 \gamma_\alpha(\omega)= D_\alpha(\omega) \frac{\lambda_\alpha(\omega)^2}{\omega}
\;.
\end{equation}
Note that we here use the definition of the bath spectral function of
Ref.~\cite{Caldeira83} without the factor $\pi / 2$, which corresponds to
the definition of Ref.~\cite{Haake85} with an additional $1 / \omega$
factor.

The linear expectation values $\bfX_\nu^\alpha$ have to scale as $1 /
 \sqrt{N_\alpha}$, because they appear in Eq.~\eqref{I} with the prefactors
$\lambda_\nu^\alpha$.
We introduce continuous functions $X_{\alpha,Q}(\omega)$ and
$X_{\alpha,P}(\omega)$ according to 
\begin{equation}
 \bfX_\nu^\alpha = \frac{\bfX_\alpha(\omega_\nu^\alpha)}{\sqrt{N}}
 = \frac{1}{\sqrt{N}} \begin{pmatrix} X_{\alpha,Q}(\omega_\nu^\alpha) \\
 X_{\alpha,P}(\omega_\nu^\alpha) \end{pmatrix} \, .
\end{equation}

Moreover, we have to separate the $N_\alpha$ diagonal terms
$\bfSig_{\nu\nu}^\alpha$ from the $N_\alpha^2$ off-diagonal terms
$\bfSig_{\nu\mu}^\alpha$ with $\nu \neq \mu$ that require an additional $1 /
N_\alpha$ prefactor for convergence in the thermodynamic limit. Hence, we define
\begin{equation}
 \bfSig_{\nu\mu}^\alpha = \bfSig_\alpha^{(1)}(\omega_\nu^\alpha) \delta_{\nu\mu}
+ \frac{1}{N_\alpha} \bfSig_\alpha^{(2)}(\omega_\nu^\alpha, \omega_\mu^\alpha)
\;,
\end{equation}
with continuous functions $\sigma_{\alpha,XY}^{(1)}(\omega)$ and
$\sigma_{\alpha,XY}^{(2)}(\omega_1, \omega_2)$ ($X, Y = Q, P$) as the matrix
entries of $\bfSig_\alpha^{(1)}(\omega)$ and $\bfSig_\alpha^{(2)}(\omega_1,
\omega_2)$.

The function $F(z)$ in the thermodynamic limit can be obtained via contour
integration with the result
\begin{equation}\begin{split}
 F(z) &= \Big( \Omega^2 + \sum_{\alpha=1}^{N_B} \int_0^\infty \rmd\omega \:
\frac{\gamma_\alpha(\omega)}{\omega} - z^2 + \sum_{\alpha=1}^{N_B} \int_0^\infty
\frac{\omega \gamma_\alpha(\omega)}{z^2 - \omega^2} \, \rmd\omega \Big)^{-1} \\
 & = \Big( \Omega^2- \sum_{\alpha=1}^{N_B} \Gamma_\alpha(\rmi 0^+) - z^2 +
\sum_{\alpha=1}^{N_B} \Gamma_\alpha(z) \Big)^{-1}
\end{split}\end{equation}
for $\Im z > 0$.
The complex functions $\Gamma_\alpha(z)$ follow from analytic continuation of
$\gamma_\alpha(\omega) = \mp (2 / \pi) \Im \Gamma_\alpha(\pm\omega + \rmi 0^+)$
into the upper half of the complex plane.

If the function $F(z)$ has no poles for $\Im z > 0$, the classical function
$u(t)$ from Eq.~\eqref{uFromFourier} is the inverse Fourier transform of a
continuous function.
We can use the Riemann-Lebesgue lemma
\begin{equation} \label{RLlemma}
 \lim_{t \to \pm\infty} \int_{-\infty}^\infty \rmd\omega \: f(\omega) \,
\rme^{\rmi \omega t} = 0
\end{equation}
valid for any integrable function $f(\omega)$ and conclude, that $u(t) \to 0$ in
the long-time limit $t \to \infty$.
In turn, poles of $F(z)$ correspond to undamped oscillations in $u(t)$, such
that the central oscillator will approach a stationary state only if isolated
modes do not exist.
The possibility of $\lim_{t \to \infty} u(t) \neq 0$, i.\,e. the existence of
isolated poles in $F(z)$, is closely connected
to a breaking of ergodicity in the sense of the mean-square of a stochastic
observable~\cite{Papoulis91,Lutz2004,Bao2005,Bao2006}.
Precise conditions for $\lim_{t \to \infty} u(t) = 0$, as well as a general discussion of equilibration and thermalization of the central oscillator, can be found in Ref.~\cite{Pagel2013}.
Throughout this work, we assume that $F(z)$ has no isolated poles, such that the
classical solutions for $t \to \infty$ approach zero, i.\,e., $\bfU(t) \to 0$.
Then, the central oscillator equilibrates and the asymptotic state is Gaussian
with the expectation values in the long-time limit $\lim_{t\to\infty} \bfX(t) =
0$ and $\lim_{t\to\infty} \bfSig(t) = \bfSig^\infty$~\cite{Pagel2013}.

\section{Nonequilibrium fluctuation relation for the oscillator
position}\label{sec:oscpos}
The results from the previous section allow us to derive a generalized
nonequilibrium fluctuation relation of the form of Eq.~\eqref{thermal0}.
For this, we determine the symmetric and the antisymmetric correlation functions
of the central oscillator position $Q$.
Their Fourier transforms are then shown to obey a generalized nonequilibrium
fluctuation relation in form of a characteristic proportionality relation.

\subsection{The symmetric correlation function}\label{sec:symmCF}
We define the symmetric correlation function of the central oscillator
position as
\begin{equation} \label{psifunc}
 \Psi(t,s) = \frac{1}{2} \Big\langle Q(t) Q(t + s) + Q(t + s) Q(t) \Big\rangle
\;.
\end{equation}
Inserting the solution for $Q(t)$ from Eq.\ (\ref{qsolution}) and performing the
thermodynamic limit $N_\alpha \to \infty$, we obtain
\begin{equation}\begin{split}
 \Psi(t,s) &= \langle Q(t) \rangle \langle Q(t + s) \rangle + \dot{u}(t)
 \dot{u}(t + s) \Sigma_{QQ}(0) + u(t) u(t + s) \Sigma_{PP}(0) \\
 &\quad + \Big( \dot{u}(t) u(t + s) + u(t) \dot{u}(t + s) \Big)
 \Sigma_{QP}(0) + \Psi^{(1)}(t, s) + \Psi^{(2)}(t, s) \:,
\end{split}\end{equation}
with the two functions 
\begin{equation}\begin{split}
 \Psi^{(1)}(t, s) &= \sum_{\alpha=1}^{N_B} \int_0^\infty \rmd\omega \: \omega \,
 \gamma_\alpha(\omega) \Bigg\{ u_R(t, \omega) u_R(t + s, \omega)
 \sigma_{\alpha,QQ}^{(1)}(\omega) \\
 &\quad + u_I(t, \omega) u_I(t + s, \omega)
 \frac{\sigma_{\alpha,PP}^{(1)}(\omega)}{\omega^2} \\
 &\quad + \Big[ u_R(t, \omega) u_I(t + s, \omega) + u_R(t + s, \omega)
 u_I(t, \omega) \Big] \frac{\sigma_{\alpha,QP}^{(1)}(\omega)}{\omega} \Bigg\}
\end{split}\end{equation}
and
\begin{equation}\begin{split}
 \Psi^{(2)}(t, s) &= \sum_{\alpha=1}^{N_B} \int_0^\infty \rmd\omega_1
\int_0^\infty
 \rmd\omega_2 \: D_\alpha(\omega_1) D_\alpha(\omega_2) \lambda_\alpha(\omega_1)
 \lambda_\alpha(\omega_2) \\
 &\quad \times \Bigg\{ u_R(t, \omega_1) u_R(t + s, \omega_2)
 \sigma_{\alpha,QQ}^{(2)}(\omega_1, \omega_2) \\
 &\qquad + u_I(t, \omega_1) u_I(t + s, \omega_2)
 \frac{\sigma_{\alpha,PP}^{(2)}(\omega_1,\omega_2)}{\omega_1 \omega_2} \\
 &\qquad + \Big[ u_R(t, \omega_1) u_I(t + s, \omega_2)
 + u_R(t + s, \omega_1) u_I(t, \omega_2) \Big]
 \frac{\sigma_{\alpha,QP}^{(2)}(\omega_1, \omega_2)}{\omega_2} \Bigg\} \;.
\end{split}\end{equation}

In the long-time limit $t \to \infty$ the terms involving $u(t)$, $\dot{u}(t)$
and $\langle Q(t) \rangle$ vanish according to our assumption of continuity of
$F(z)$.
For the remaining terms $\Psi^{(1)}(t, s)$ and $\Psi^{(2)}(t, s)$ we rewrite the
partial Fourier transform of Eq.\ (\ref{partfourier}) as
${u}(t + s, \omega) = \rme^{\rmi \omega s} \left[ {u}(t,\omega) - \int_0^s
\rmd\tau \: u(t + \tau) \, \rme^{-\rmi \omega \tau} \right]$.
Since $u(t)$ vanishes at long times, the partial Fourier transform ${u}(t + s,
\omega)$ behaves asymptotically as
\begin{equation} \label{u_as}
 u_{as}(t + s, \omega) \simeq \rme^{\rmi \omega (t + s)} {u}(\omega) \;,
\end{equation}
where 
\begin{equation} \label{FTu}
 u(\omega) = \int_0^\infty \rmd\tau \: u(\tau) \, \rme^{-\rmi \omega \tau}
\end{equation}
is the full Fourier transform of the function $u(t)$
\footnote{We use the same symbol $u$ for the function and its Fourier
transform for ease of readability. Time arguments are denoted as $t,\tau$ or
$s$, while frequency arguments are denoted by $\omega$.}.
Using this asymptotic behaviour in the expressions for $\Psi^{(1)}(t, s)$ and
$\Psi^{(2)}(t, s)$ we see that the off-diagonal term $\Psi^{(2)}(t, s)$ contains
only oscillatory 
terms in the two frequencies $\omega_1$ and $\omega_2$.
If we recall the Riemann-Lebesgue lemma, Eq.\ \eqref{RLlemma}, we conclude, that
$\Psi^{(2)}(t, s)$ vanishes in the long-time limit.
Following the same line of reasoning we find that the only non-zero term in
the limit $t \to \infty$ comes from $\Psi^{(1)}(t, s)$ and involves
$|u(\omega)|^2$ while the arising oscillating terms vanish. 
In particular,
\begin{equation} \label{symcor}
 \Psi(s) \equiv \lim_{t\to\infty} \Psi(t, s) = \sum_{\alpha=1}^{N_B}
\int_0^\infty
 \rmd\omega \: \gamma_\alpha(\omega) \, |u(\omega)|^2
 \frac{\mathcal{E}_\alpha(\omega)}{\omega} \cos \omega s \;,
\end{equation}
where 
\begin{equation} \label{freresol}
 \mathcal{E}_\alpha(\omega) = \frac{1}{2} \Big( \omega^2
 \sigma_{\alpha,QQ}^{(1)}(\omega) + \sigma_{\alpha,PP}^{(1)}(\omega) \Big)
\end{equation}
denotes the frequency-resolved energy distribution functions of the initial bath
states.

We next Fourier transform Eq.\ (\ref{symcor}) and obtain
\begin{equation} \label{finalsymcor}
 \Psi(\omega) = \int_{-\infty}^\infty \rmd s \: \rme^{\rmi \omega s} \Psi(s)
 = \pi \sum_{\alpha=1}^{N_B} \gamma_\alpha(\omega) \, |u(\omega)|^2
 \frac{\mathcal{E}_\alpha(\omega)}{\omega} \;.
\end{equation}

\subsection{The antisymmetric correlation function}
The antisymmetric correlation function of the oscillator position $Q$ is given
by 
\begin{equation}
  \Phi(t, s) = \frac{1}{\rmi} \Big\langle Q(t) Q(t + s) - Q(t + s) Q(t)
\Big\rangle \:.
\end{equation}
Inserting the solution $Q(t)$ of Eq.\ (\ref{qsolution}), using the property 
$\langle [Q_\nu^\alpha(0), P_\mu^\alpha(0)] \rangle = \rmi \delta_{\nu\mu}$ 
and performing the thermodynamic limit $N_\alpha \to \infty$, we obtain
\begin{equation}\begin{split}
 \Phi(t, s) &= \dot{u}(t) u(t + s) - u(t) \dot{u}(t + s) \\
 &\quad + \sum_{\alpha=1}^{N_B} \int_0^\infty \rmd\omega \:
\gamma_\alpha(\omega)
 \Big[ u_R(t, \omega) u_I(t + s, \omega) - u_R(t + s, \omega) u_I(t, \omega)
 \Big] \;, 
\end{split}\end{equation}
which is independent of the initial bath preparation as expected
\cite{Weiss2008}.

Similar to the calculation of the symmetric correlation function, we obtain for
the antisymmetric response function in the long-time limit
\begin{equation}
 \Phi(s) \equiv \lim_{t\to\infty} \Phi(t, s) = \sum_{\alpha=1}^{N_B}
\int_0^\infty
 \rmd\omega \: \gamma_\alpha(\omega) \, |u(\omega)|^2 \sin \omega s \:.
\end{equation}
Its Fourier transform readily follows as
\begin{equation} \label{finalasymcor}
 \Phi(\omega) = \int_{-\infty}^\infty \rmd s \: \rme^{\rmi \omega s} \Phi(s)
 = \rmi \pi \sum_{\alpha=1}^{N_B} \gamma_\alpha(\omega) \, |u(\omega)|^2 \:.
\end{equation}

\subsection{The generalized nonequilibrium fluctuation
relation}\label{sec:nfrpos}
To formulate the general nonequilibrium fluctuation relation, we compare 
Eqs.\ (\ref{finalsymcor}) and (\ref{finalasymcor}) and obtain for general
initial preparations and an arbitrary number $N_B$ of independent harmonic
baths the relation  
\begin{equation} \label{neqfr}
 \Psi(\omega) = \frac{1}{\rmi} \frac{\sum_{\alpha=1}^{N_B}
 \gamma_{\alpha}(\omega) \mathcal{E}_{\alpha}(\omega)}{\omega
 \sum_{\alpha=1}^{N_B} \gamma_{\alpha}(\omega)} \Phi(\omega) \:.
\end{equation}
This is one major result of the present work and illustrates that the relation
is crucially determined by the frequency-resolved energy
distributions ${\mathcal{E}}_{\alpha}(\omega)$ of the initial bath states
defined in Eq.\ (\ref{freresol}) and the bath spectral functions
$\gamma_{\alpha}(\omega)$ given in Eq.\ (\ref{specfunc}).
A comparison with the thermal fluctuation-dissipation theorem in
Eq.\ \eqref{thermal0} shows that in the considered nonthermal situation we have
to exchange the thermal energy distribution
\begin{equation}
 \mathcal{E}_{\text{th}}(\omega, T) = \frac{\omega}{2} \coth \frac{\omega}{2 T}
\end{equation}
with the average of the individual energy distributions of the baths weighted
with their spectral functions.

In the case when all baths are initially distributed thermally at the
same temperature $T$ according the thermal equilibrium Bose-Einstein
distribution function, we have $\mathcal{E}_\alpha(\omega) =
\mathcal{E}_{\text{th}}(\omega, T)$ for all $\alpha = 1, \dots, N_B$.
This reproduces the
equilibrium fluctuation-dissipation theorem Eq.~\eqref{thermal0} 
\cite{Weiss2008}.

A natural question then is under which initial bath preparations the central
oscillator thermalizes, i.\,e., reaches a stationary state which is thermally
distributed with a given temperature $T$.
By comparing Eqs.\ (\ref{neqfr}) and (\ref{thermal0}), we obtain the condition
\begin{equation}
 \frac{\sum_\alpha \gamma_\alpha(\omega) \mathcal{E}_\alpha(\omega)}
 {\sum_\alpha \gamma_\alpha(\omega)} = \mathcal{E}_{\text{th}}(\omega, T)
\end{equation}
for which the fluctuations of the central oscillator for $t \to \infty$ are
thermal.
This condition certainly is satisfied whenever all baths are thermal and have 
equal temperature, but can also be satisfied for other nonthermal
initial bath preparations. In turn, if this condition is satisfied,
the quantity
\begin{equation} 
 T^{-1} = \frac{2}{\omega} \text{arcoth}\bigg( \frac{2}{\omega}
\frac{\sum_{\alpha} \gamma_\alpha(\omega)
{\mathcal{E}}_\alpha(\omega)}{\sum_\alpha \gamma_\alpha(\omega)} \bigg)
\end{equation}
is a constant, i.\,e., independent of $\omega$.
It is then tempting to understand this quantity as an
``effective'' temperature characterizing the general initial bath preparation.
However, the above condition does not guarantee true thermalization of the
central oscillator, which is essential for a meaningful notion of temperature.
For a more detailed discussion of this question, see Ref.~\cite{Pagel2013}.

\section{Generating function for the position operator of the oscillator}
\label{sec:genfuncpos}
In this section we show that the dissipative oscillator model allows us to
study the connection between transient and steady state fluctuation relations.
We calculate the generating function for the central oscillator position
operator and show that it fullfills a Gallavotti-Cohen symmetry
relation~\cite{Gallavotti95} valid for arbitrary times and a Gaussian initial
state of the central oscillator.
This additional Gaussian assumption is not necessary in the long-time limit and
we obtain an exact result for the steady state fluctuation relation.

We define the generating function for the position operator according to 
\begin{equation}
 Z_Q(\xi, t) = \langle \rme^{\rmi \xi Q(t)} \rangle \;.
\end{equation}
With that, all the cumulants $\langle\langle Q^n(t) \rangle\rangle$ of the
position 
operator follow by performing the respective derivative,
\begin{equation}
 \langle\langle Q^n(t) \rangle\rangle = \frac{\partial^n \ln Z_Q(\xi, t)}
 {\partial (\rmi \xi)^n} \bigg|_{\xi=0} \;.
\end{equation}
For instance, we have $\langle\langle Q(t) \rangle\rangle = \langle Q(t)
\rangle$ and $\langle\langle Q^2(t) \rangle\rangle = \Sigma_{QQ}(t)$.

It is convenient to represent the generating function in terms of the Wigner
function of the central oscillator
\begin{equation}
 W_S(q, p, t) = \frac{1}{2 \pi} \int_{-\infty}^\infty \rmd s \:
 \Big\langle q + \frac{s}{2} \Big| \rho_S(t) \Big| q - \frac{s}{2} \Big\rangle
 \, \rme^{-\rmi p s} \;,
\end{equation}
such that
\begin{equation}
 Z_Q(\xi, t) = \int_{\mathbb{R}^2} \rmd\bfx \: W_S(\bfx, t)\,\rme^{\rmi \xi q}
\;,
\end{equation}
where we write $W_S(\bfx, t) = W_S(q, p, t)$ with $\bfx = (q, p)^T$ and 
$\rmd\bfx = \rmd q \, \rmd p$ for abbreviation.
The Wigner function $W_S(\bfx, t)$ at time $t \geq 0$ can be obtained from the 
propagating function $J_W(\bfx, \bar{\bfx}, t) = J_W(q, p, \bar{q}, \bar{p},
t)$ 
in Wigner representation, that is defined by the relation
\begin{equation}
 W_S(\bfx, t) = \int_{\mathbb{R}^2} \rmd\bar{\bfx} \: J_W(\bfx, \bar{\bfx}, t)
 W_S(\bar{\bfx}, 0) \;,
\end{equation}
and can be evaluated to~\cite{Pagel2013}
\begin{equation}
 J_W(\bfx, \bar{\bfx}, t) = \frac{\exp \left\{ -\frac{1}{2} \left[ \bfx -
\bfU(t) \bar{\bfx} - \bfI(t) \right] \cdot \bfC^{-1}(t) \left[ \bfx - \bfU(t)
\bar{\bfx} - \bfI(t) \right] \right\}}{2 \pi \sqrt{\det \bfC(t)}} \;,
\end{equation}
with $\bfU(t)$, $\bfI(t)$, and $\bfC(t)$ from Eqs.~\eqref{U}, \eqref{I}, and
\eqref{C}.
Performing the Gaussian integral over $\bfx$ we obtain
\begin{equation} \label{ZQ_finiteT}
 Z_Q(\xi, t) = \int_{\mathbb{R}^2} \rmd\bar{\bfx} \: W_S(\bar{\bfx}, 0) \, \exp
 \left\{ -\frac{\xi^2}{2} \bfe_1 \cdot \bfC(t) \bfe_1 + \rmi \xi [\bfU(t)
\bar{\bfx} + \bfI(t)] \cdot \bfe_1 \right\}
\end{equation}
where $\bfe_1 = (1, 0)^T$.

In the long-time limit $t \to \infty$, where $\bfU(t) \to 0$ according to our
assumption of continuity of $F(z)$, the integration in Eq.~\eqref{ZQ_finiteT}
evaluates to one because the initial Wigner function is normalized.
We then obtain
\begin{equation}
 Z_Q^\infty(\xi) \equiv \lim_{t \to \infty} Z_Q(\xi, t) = \exp \bigg\{
-\frac{\xi^2}{2} \Sigma_{QQ}^\infty \bigg\}
\end{equation}
with $\Sigma_{QQ}^\infty = \lim_{t\to\infty} \Sigma_{QQ}(t)$.
The results obeys the symmetry $Z_Q^\infty(\xi) = Z_Q^\infty(-\xi)$.

For finite times, we can restrict ourselves to Gaussian initial states of the
central oscillator,
\begin{equation}
 W_S(\bar{\bfx}, 0) = \frac{\exp \left\{ -\frac{1}{2} [\bar{\bfx} - \bfX(0)]
 \cdot \bfSig^{-1}(0) [\bar{\bfx} - \bfX(0)] \right\}}
 {2 \pi \sqrt{\det \bfSig(0)}} \;,
\end{equation}
and obtain
\begin{equation}\begin{split}
 Z_Q(\xi, t) &= \exp \left\{ -\frac{\xi^2}{2} \bfe_1 \cdot \bfSig(t) \bfe_1
 + \rmi \xi \bfX(t) \cdot \bfe_1 \right\} \\
 &= \exp \left\{ -\frac{\xi^2}{2} \Sigma_{QQ}(t) + \rmi \xi
 \langle Q(t) \rangle \right\} \;.
\end{split}\end{equation}
In order to see when the Gallavotti-Cohen relation is fullfilled, we calculate 
\begin{equation}
 Z_Q(-\xi + \rmi A, t) = \exp \left\{ -\frac{\xi^2}{2} \Sigma_{QQ}(t) + \rmi \xi
 [A \Sigma_{QQ}(t) - \langle Q(t) \rangle] + \frac{A}{2} [A \Sigma_{QQ}(t)
 - 2 \langle Q(t) \rangle] \right\} \;.
\end{equation}
Hence, the relation $Z_Q(-\xi + \rmi A, t) = Z_Q(\xi, t)$ is fullfilled at any
arbitrary time $t$, if
\begin{equation}
 A \equiv A(t) = 2 \frac{\langle Q(t) \rangle}{\Sigma_{QQ}(t)} \;.
\end{equation}
This implies that the oscillator fluctuates symmetrically around its momentary
position average $\langle Q(t) \rangle$ since
$Z_{Q-\langle Q \rangle}(-\xi, t) = Z_{Q-\langle Q \rangle}(\xi, t)$. On 
the other
hand, however, the symmetry point for the generating function of the position
operator,
which in the stationary state is $\xi=0$, is shifted by the momentary position
expectation 
value scaled by the momentary position variance, i.\,e. $Z_Q(-\xi+\rmi A/2, t) =
Z_Q(\xi+\rmi A/2, t)$.
Note that this relation holds in general and also when the central oscillator
has not yet reached its equilibrium state.
This transient fluctuation relation is linked with the steady state fluctuation
relation from above by realizing that $\lim_{t\to\infty} A(t) = 0$.

\section{Quantum mechanical energy transfer between nonequilibrium baths}
\label{sec.heat}
We now study the quantum mechanical transfer of energy between
nonequilibrium baths. To keep the discussion simple, we
concentrate on the case of the energy transfer between two baths, i.\,e., $N_B=2$,
and denote them as left ($\alpha=l$) and right ($\alpha=r$) reservoir. 
In particular, we are interested in the form of the expectation value of the
energy current operator which can be defined for instance
for the left junction
according to
\cite{Talkner1990,Segal2003,Dhar2006,Saito2007,Saito2007E,Agarwalla2012}
\begin{equation} \label{HeatTransf}
 I(t) = -\frac{\rmd H_B^l(t)}{\rmd t} = \frac{1}{2} \sum_{\nu=1}^{N_l}
\lambda_\nu^l \big\{ P_\nu^l(t), Q(t) \big\}
\end{equation}
with the anticommutator defined as $\{A, B\} = AB + BA$.

For the calculation of the expectation value $\langle I(t) \rangle$ we need the
solutions of the Heisenberg equations of motion for the left bath operators,
\begin{subequations} \begin{align}\label{bathOpsQ}
 Q_\nu^l(t) &= \cos \omega_\nu^l t \, Q_\nu^l(0) + \frac{\sin \omega_\nu^l
t}{\omega_\nu^l} \, P_\nu^l(0) - \lambda_\nu^l \int_0^t \rmd\tau \: \frac{\sin
\omega_\nu^l (t - \tau)}{\omega_\nu^l} \, Q(\tau) \;, \\\label{bathOpsP}
 P_\nu^l(t) &= \dot{Q}_\nu^l(t) \;.
\end{align}\end{subequations}
We insert these equations and the solution Eq.~\eqref{qsolution} for $Q(t)$ into
Eq.~\eqref{HeatTransf} and perform the thermodynamic limit to obtain $\langle
I(t)
\rangle = \langle I_1(t) \rangle + \langle I_2(t) \rangle + \langle I_3(t)
\rangle$ with
\begin{subequations}\begin{align}
 \langle I_1(t) \rangle &= \int_0^\infty \rmd\omega \: D_l(\omega)
\lambda_l(\omega) \omega \sin \omega t \, {X}_{l,Q}(\omega) \langle Q(t)
\rangle + \langle I_1^{(1)}(t) \rangle + \langle I_1^{(2)}(t) \rangle \;, \\
 \langle I_2(t) \rangle &= -\int_0^\infty \rmd\omega \: D_l(\omega)
\lambda_l(\omega) \cos \omega t \, {X}_{l,P}(\omega) \langle Q(t) \rangle
+ \langle I_2^{(1)}(t) \rangle + \langle I_2^{(2)}(t) \rangle \;, \\ 
 \langle I_3(t) \rangle &= \int_0^\infty \rmd\omega \: \omega \gamma_l(\omega)
\int_0^t \rmd\tau \: \cos \omega \tau \, \Psi(t, \tau) \;.
\end{align}\end{subequations}
In these equations, $\Psi(t,\tau)$ is the symmetric position autocorrelation
function given in Eq.\ (\ref{psifunc}) and the diagonal and non-diagonal
contributions to $\langle I_1(t) \rangle$ and $\langle I_2(t) \rangle$ are
\begin{subequations}\begin{align}
 \langle I_1^{(1)}(t) \rangle &= -\int_0^\infty \rmd\omega \: \omega^2
\gamma_l(\omega) \sin \omega t \Big[ {u}_R(t, \omega)
{\sigma}_{l,QQ}^{(1)}(\omega) + \frac{{u}_I(t, \omega)}{\omega}
{\sigma}_{l,QP}^{(1)}(\omega) \Big] \;, \\
 \langle I_1^{(2)}(t) \rangle &= -\int_0^\infty \rmd\omega_1 \int_0^\infty
\rmd\omega_2 \: D_l(\omega_1) D_l(\omega_2) \lambda_l(\omega_1)
\lambda_l(\omega_2) \omega_1 \sin \omega_1 t \nonumber\\
 &\qquad \times \Big[ {u}_R(t, \omega_2)
{\sigma}_{l,QQ}^{(2)}(\omega_1, \omega_2) + \frac{{u}_I(t,
\omega_2)}{\omega_2} {\sigma}_{l,QP}^{(2)}(\omega_1, \omega_2) \Big] \;,
\end{align}\end{subequations}
and
\begin{subequations}\begin{align}
 \langle I_2^{(1)}(t) \rangle &= \int_0^\infty \rmd\omega \: \omega
\gamma_l(\omega) \Big[ \cos \omega t \, {u}_I(t, \omega)
{\sigma}_{l,QP}^{(1)}(\omega) + \cos \omega t \, \frac{{u}_I(t,
\omega)}{\omega} {\sigma}_{l,PP}^{(1)}(\omega) \Big] \;, \\
 \langle I_2^{(2)}(t) \rangle &= \int_0^\infty \rmd\omega_1 \int_0^\infty
\rmd\omega_2 \: D_l(\omega_1) D_l(\omega_2) \lambda_l(\omega_1)
\lambda_l(\omega_2) \nonumber\\
 &\qquad \times \Big[ \cos \omega_2 t \, {u}_R(t, \omega_1)
{\sigma}_{l,QP}^{(2)}(\omega_1, \omega_2) + \cos \omega_1 t \,
\frac{{u}_I(t, \omega_2)}{\omega_2} {\sigma}_{l,PP}^{(2)}(\omega_1,
\omega_2) \Big] \;.
\end{align}\end{subequations}

To perform the long-time limit, we follow the line of reasoning of
Sec.~\ref{sec:symmCF}.
The terms containing the linear expectation value $\langle Q(t) \rangle$ vanish.
The off-diagonal terms $\langle I_1^{(2)}(t) \rangle$ and $\langle I_2^{(2)}(t)
\rangle$ contain oscillatory terms in the two frequencies $\omega_1$ and
$\omega_2$ only, such that $\langle I_1^{(2)}(t) \rangle, \langle I_2^{(2)}(t)
\rangle \to 0$ for $t \to \infty$.
The remaining diagonal terms can be simplified algebraically using the
asymptotic behaviours Eq.~\eqref{u_as} of the partial Fourier transform $u(t,
\omega)$ and Eq.~\eqref{symcor} of the symmetric correlation function $\Psi(t,
\tau)$ and by applying the Riemann-Lebesgue lemma Eq.~\eqref{RLlemma}.
We finally obtain the expectation value of the energy current from the left
reservoir to the central oscillator in the long-time limit as 
\begin{equation}
 I_\infty \equiv \lim_{t \to \infty} \langle I(t) \rangle = -\int_0^\infty
\rmd\omega \: \gamma_l(\omega) \Big[ {u}_I(\omega)
{\mathcal{E}}_l(\omega) + \frac{\pi}{2} \sum_{\alpha=l,r} \gamma_\alpha(\omega)
|{u}(\omega)|^2 {\mathcal{E}}_\alpha(\omega) \Big] \;.
\end{equation}
We can rewrite this expression into the final form
\begin{equation} \label{HeatTransp_final}
 I_\infty = \frac{\pi}{2} \int_0^\infty \rmd\omega \: \gamma_l(\omega)
\gamma_r(\omega) |{u}(\omega)|^2 \Big[ {\mathcal{E}}_l(\omega) -
{\mathcal{E}}_r(\omega) \Big]
\end{equation}
by using, that the Fourier transform $u(\omega)$ in Eq.~\eqref{FTu} is the
inverse of the Fourier transform in Eq.~\eqref{uFromFourier}, such that
$u_I(\omega) = -\Im F(\omega + \rmi 0^+) = -(\pi / 2) [\gamma_l(\omega) +
\gamma_r(\omega)] |u(\omega)|^2$.

Expression~\eqref{HeatTransp_final} generalizes Eq.~(4.2) of Ref.\
\cite{Dhar2006} and reproduces it for the special case of thermal baths.
Obviously, the asymptotic energy current vanishes exactly, if
$\gamma_r(\omega) = 0$ for only one bath, or if $\mathcal{E}_r(\omega) =
\mathcal{E}_l(\omega)$ for equal bath preparations.

For two thermal baths with $\mathcal{E}_\alpha(\omega) =
\mathcal{E}_{\text{th}}(\omega, T_\alpha)$, and $T_l = T_r + \Delta T$ where
$\Delta T
\ll T_l$, we can expand the energy distribution function as
\begin{equation}
{\mathcal{E}}_l(\omega) = {\mathcal{E}}_r(\omega) + \Big[
\frac{\omega}{2 T_r} \sinh^{-1} \frac{\omega}{2 T_r} \Big]^2 \Delta T +
\mathcal{O}\left(\Delta T^2\right) \;,
\end{equation}
where $\sinh^{-1} x = 1 / \sinh x$.
Thus, we obtain the linear response result
\begin{equation}
 I_\infty^{(\text{lin})} =  \Delta T \frac{\pi}{2} \int_0^\infty \rmd\omega \:
\gamma_l(\omega) \gamma_r(\omega) |{u}(\omega)|^2 \frac{\omega^2}{4 T_r^2}
\sinh^{-2} \frac{\omega}{2 T_r} + \mathcal{O}\left(\Delta T^2\right)
\end{equation}
growing linearly with the difference $\Delta T$ of the temperatures of the left
and right bath.

\section{Nonequilibrium fluctuations of the transferred energy}
\label{sec.heatflucs}
In this section, we consider the energy which is transferred from one
bath (say, the left) to the central oscillator in presence of the second
bath (say, the right). Moreover, we are interested in the fluctuations of the
transferred energy. We note in passing that we use the more general term of
``energy'' instead of ``heat'' since the definition of heat in the strict sense
requires purely thermal environments.

The energy that is transferred from the left bath to the rest of the system
until time $t$ is obtained from the difference of the energy of the left bath
between times $t$ and $0$. This involves the measurement of the
observable $H_B^l$ at two different times.
Following the idea of two-time quantum measurements, the corresponding
generating function can be written as~\cite{Saito2007,Saito2007E,Agarwalla2012}
\begin{equation}
 Z(\xi, t) = \big\langle \rme^{\rmi \xi H_B^l} \rme^{-\rmi \xi H_B^l(t)}
\big\rangle' \;,
\end{equation}
where the prime indicates that the expectation value has to be taken with
respect to the projected density matrix
\begin{equation}
 \rho'(0) = \sum_a | \phi_a \rangle \langle \phi_a | \rho(0) | \phi_a
\rangle \langle \phi_a | \;.
\end{equation}
Here $| \phi_a \rangle$ is an eigenstate of the operator $H_B^l$, i.\,e., 
$H_B^l | \phi_a \rangle = a | \phi_a \rangle$.
Writing the generating function as a series in powers of $\rmi \xi$, we
obtain~\cite{Saito2007,Agarwalla2012}
\begin{equation}\label{lnZ_momentsW}
 \ln Z(\xi, t) = \sum_{n=1}^\infty \frac{(\rmi \xi)^n}{n!} \langle\langle
W^n(t) \rangle\rangle \;,
\end{equation}
where $\langle\langle W^n(t) \rangle\rangle$ denotes the $n$th order
cumulant of the operator
\begin{equation} \label{W}
 W(t) = \int_0^t \rmd\tau \: I(\tau) = H_B^l(0) - H_B^l(t) \;.
\end{equation}
In the following, we calculate the moments of the energy transfer operator
$W(t)$ entering Eq.~\eqref{lnZ_momentsW}.
In particular, we are interested in the long-time limit of these quantities.

\subsection{The first moment}
Using Eqs.~\eqref{bathOpsQ} and~\eqref{bathOpsP} the linear expectation value
of the energy transfer operator follows as
\begin{equation}\begin{split}
 \langle W(t) \rangle &= -\frac{1}{2} \sum_{\nu=1}^{N_l} \bigg\langle
\lambda_\nu^l \int_0^t \rmd\tau \: \Big( \omega_\nu^l 
\sin \omega_\nu^l \tau \big\{ Q_\nu^l(0), Q(\tau) \big\}
 - \cos \omega_\nu^l \tau \, 
\big\{ P_\nu^l (0), Q(\tau) \big\} \Big) \\
 &\quad + \frac{(\lambda_\nu^l)^2}{2} \int_0^t \int_0^t \rmd\tau \,
\rmd\bar{\tau} \:
\cos \omega_\nu^l (\tau - \bar{\tau}) \{ Q(\tau), Q(\bar{\tau})\} 
\bigg\rangle \;.
\end{split}\end{equation}
In the long-time limit $t \to \infty$, we expect from the definition
in Eq.~\eqref{W} and from the result $\langle I(t) \rangle \to I_\infty$ of the
last section that $\langle W(t) \rangle$ grows linearly with time.
It is thus useful to consider $\langle W(t) \rangle / t$ instead of $\langle
W(t) \rangle$.

The explicit calculation of $\langle W(t) \rangle / t$ in the long-time limit is
achieved by inserting the solution $Q(t)$ from Eq.~\eqref{qsolution},
performing the thermodynamic limit according to Sec.~\ref{sec:tdLimit}, and
analytically carrying out the remaining time integrations.
The result is
\begin{equation}\begin{split}
 \lim_{t \to \infty} \frac{1}{t} \langle W(t) \rangle &= -\bigg\{ \int_0^\infty
\rmd\omega \: \gamma_l(\omega) {u}_I(\omega)
{\mathcal{E}}_l(\omega) + \frac{\pi}{2} \sum_{\alpha=l,r} \int_0^\infty
\rmd\omega \: \gamma_l(\omega) \gamma_\alpha(\omega) |{u}(\omega)|^2
{\mathcal{E}}_\alpha(\omega) \bigg\} \\
 &= \frac{\pi}{2} \int_0^\infty \rmd\omega \: \gamma_l(\omega) \gamma_r(\omega)
|{u}(\omega)|^2 \left[{\mathcal{E}}_l(\omega) -
{\mathcal{E}}_r(\omega) \right] \;.
\end{split}\end{equation}
As expected, this expression coincides with the expectation value of the energy 
current operator given in Eq.\ (\ref{HeatTransp_final}).

\subsection{The second moment}
The second moment of the energy transfer operator is
\begin{equation}\begin{split}
 \langle W^2(t) \rangle &= \bigg\langle \bigg[ \sum_{\nu=1}^{N_l}
\bigg\{ \frac{\lambda_\nu^l}{2} \int_0^t \rmd\tau \: \Big( \omega_\nu^l \sin
\omega_\nu^l \tau \{Q_\nu^l(0), Q(\tau)\}
 - \cos \omega_\nu^l \tau \{P_\nu^l(0), Q(\tau)\} \Big) \\
 &\quad + \frac{(\lambda_\nu^l)^2}{4} \int_0^t \int_0^t \rmd\tau \,
\rmd\bar{\tau} \:
\cos \omega_\nu^l (\tau - \bar{\tau}) \{Q(\tau), Q(\bar{\tau})\}
\bigg\} \bigg]^2 \bigg\rangle \;.
\end{split}\end{equation}
Expanding the square yields a sum of terms containing expectation values of
products of four operators.
We may reorder the operator products using the commutators
\begin{equation}
 [Q_\nu^l(0), Q(t)] = -\rmi \frac{\lambda_\nu^l}{\omega_\nu^l} u_I(t,
\omega_\nu^l)
\;, \qquad
 [P_\nu^l(0), Q(t)] = \rmi \lambda_\nu^l u_R(t, \omega_\nu^l) \;.
\end{equation}

A general expectation value of a product of four operators can be ascribed to a
sum of products of expectation values of one or two operators for 
Gaussian states. The assumption of a Gaussian bath state is justified in the 
thermodynamic and long-time limit on general
grounds~\cite{Cramer2008,Cramer2010}. 
In Ref.~\cite{Pagel2013} it is shown that the state of the central
oscillator becomes Gaussian for $t \to \infty$, independent of its initial
preparation, if the classical solution $u(t)$ vanishes asymptotically---the 
situation of interest here.

For an explicit result, we insert the solution $Q(t)$ from
Eq.~\eqref{qsolution}, perform the thermodynamic limit, use the results for the
position correlation functions from Sec.~\ref{sec:oscpos}, and carry out the
remaining time integrals in the long-time limit.
The result for the second order cumulant then reads
\begin{equation} \label{secmom} \begin{split}
 \lim_{t \to \infty} \frac{\langle\langle W^2(t) \rangle\rangle}{t}
&= \frac{\pi^3}{2} \int_0^\infty \rmd\omega \: \gamma_l^2(\omega)
\gamma_r^2(\omega) |u(\omega)|^4 \Big( {\mathcal{E}}_l(\omega) -
{\mathcal{E}}_r(\omega) \Big)^2 \\
 &\quad + \frac{\pi}{2} \int_0^\infty \rmd\omega \: \gamma_l(\omega)
\gamma_r(\omega) |{u}(\omega)|^2 \Big( 2 {\mathcal{E}}_l(\omega)
{\mathcal{E}}_r(\omega) - \frac{\omega^2}{2} \Big) \;.
\end{split} \end{equation}
This expression generalizes the result for the second moment given in Eq.\
(9) in Ref.\ \cite{Saito2007,Saito2007E}. Eq.\ (\ref{secmom}) reduces to this
equation for
the special case of thermal baths with $\mathcal{E}_\alpha(\omega) =
\mathcal{E}_{\text{th}}(\omega, T_\alpha) = \omega f_\alpha(\omega) + \omega /
2$ which
then lead to the expressions 
$f_\alpha(\pm \omega) = 1 / [\exp(\pm\omega / T_\alpha) - 1]$ in Ref.\
\cite{Saito2007,Saito2007E}.

\subsection{The generating function for the energy transfer}
We have seen that the well-known results \cite{Saito2007,Agarwalla2012} for the
first and
second moment of the heat transfer operator for the special case of thermal 
baths are well reproduced by our more general results. The generalization
follows by the corresponding replacements of the thermal distribution functions
of the baths by the general initial distributions. Hence we can now
follow the same line of reasoning and generalize the steady state
expression of the
cumulant generating function for the heat transfer given in Eq.\ (8) in
Ref.\ \cite{Saito2007,Saito2007E} with the result
\begin{equation}\label{G_steady}\begin{split}
 G(\xi) &\equiv \lim_{t \to \infty} \frac{\ln Z(\xi, t)}{t} \\
 &= -\frac{1}{2 \pi} \int_0^\infty \rmd\omega \: \ln \bigg\{ 1 +
 \pi^2 \gamma_l(\omega) \gamma_r(\omega) \frac{|u(\omega)|^2}{\omega^2} \\
 &\quad \times \bigg[ \Big( 2 \mathcal{E}_l(\omega) \mathcal{E}_r(\omega) -
\frac{\omega^2}{2} \Big) \big( 1 - \cos \xi \omega \big) - \rmi \omega \Big(
\mathcal{E}_l(\omega) - \mathcal{E}_r(\omega) \Big) \sin \xi \omega \bigg]
\bigg\} \;.
\end{split}\end{equation}
We observe that $G(\xi)$ fullfills the symmetry relation
\begin{equation}\label{GCsym_steady}
 G(\xi) = G(-\xi + \rmi A) \;,
\end{equation}
where $A = \beta_r - \beta_l$ with 
\begin{equation} 
 \beta_\alpha = \frac{2}{\omega} \text{arcoth}\bigg(
\frac{2{\mathcal{E}}_\alpha(\omega)}{\omega}\bigg) \;.
\end{equation}
Since the constants $\beta_\alpha$ should be independent of $\omega$ the
existence of the symmetry~\eqref{GCsym_steady} implies a condition on the
initial bath preparation.
In particular, the energy distribution functions should be thermal, i.\,e.
$\mathcal{E}_\alpha(\omega) = \mathcal{E}_{\mathrm{th}}(\omega, T_\alpha)$.
Note that this is a condition on the combination $\mathcal{E}_\alpha(\omega)$ of
the initial bath variances $\sigma_{\alpha,QQ}^{(1)}(\omega)$ and
$\sigma_{\alpha,PP}^{(1)}(\omega)$, not on the individual functions [see
Eq.~\eqref{freresol}].
It can be fullfilled for nonthermal bath preparations as well~\cite{Pagel2013}.

From the relation~\eqref{GCsym_steady} it follows that the probability
distribution of the transferred energy,
\begin{equation}
 P(W) = \int_{-\infty}^\infty \frac{\rmd \xi}{2 \pi} \, \lim_{t \to \infty}
Z(\xi, t) \, \rme^{-\rmi \xi W} \;,
\end{equation}
fullfills the steady state fluctuation theorem
\begin{equation}\label{SSFT}
 P(W) = \rme^{A W} P(-W) \;.
\end{equation}
We remark that the exchange fluctuation relation~\eqref{SSFT} can only be proven
rigorously when the initial preparation is indeed free of correlations and also
the interaction of the system and the bath is switched off at some final time\ 
\cite{Campisi11,Campisi2010,Campisi2011}.
The role of initial system-bath correlations
for the nonequilibrium fluctuation relations is still an open problem.
According to that the result in Eq.~\eqref{G_steady} and the corresponding
symmetry relation~\eqref{GCsym_steady} are formulated and valid in the
long-time limit only. For transient times $t < \infty$, we expect additional
contributions to the steady state fluctuation theorem
in Eq.~\eqref{SSFT}~\cite{Campisi11,Campisi2010,Campisi2011}.

\section{Summary} \label{sec:concl}
Most studies related to fluctuation relations so far consider
the special case when the systems are initially in thermal equilibrium, but do
not restrict their analyses to a specific model. 
In the present study, we give up the assumption of initial thermal states and
allow for nonthermal bath preparations. 
The price we have to pay for this generalization is the restriction to
an analytically solvable model for which we
obtain exact generalized nonequilibrium fluctuation relations. 
On the one hand, we can give the explicit expressions for the symmetric and
antisymmetric autocorrelation functions of the central oscillator position.
Then,
a generalized nonequilibrium fluctuation relation follows which only involves 
the bath spectral functions and the frequency-resolved energy
distribution of the initial bath states. The general expression also
contains the special case of a single thermal bath and coincides with the
well-known equilibrium fluctuation-dissipation theorem. Moreover, we discuss
the conditions under which the generating function of the oscillator position
fullfills a Gallavotti-Cohen relation at arbitrary times.
This relation reflects the fact that
the oscillator position fluctuates symmetrically around its
momentary average position.
On the other hand, we have elucidated the quantum mechanical energy transfer
through the
central oscillator by calculating the time-dependent energy 
current and the second moment of the current fluctuations.
Based on this result we generalize the cumulant generating function for energy
transfer, which is well-known for thermal baths, to the nonthermal situation.

\section*{Acknowledgements}
We acknowledge support by the DFG through SFB 652 (project B5), SFB 925
(project C8), and SFB 668 (project B16).

\appendix

\section{Correlation functions of the fluctuating forces}\label{CF_flukForce}
We here give the correlation functions of the fluctuating operator-valued
forces in the quantum Langevin equation~\eqref{QLE}. By this, we illustrate
some subtleties of the nonthermal initial bath preparations with respect to
stationarity and ergodicity. 

The statistics of the operator-valued noise forces $\eta_\alpha(t)$ is
determined through their respective moments and explicitly depends on the
initial preparation of the baths.
For the common thermal bath preparation the linear expectation values
\begin{equation}
 \langle \eta_\alpha(t) \rangle = \sum_{\nu=1}^{N_\alpha} \lambda_\nu^\alpha
\Big( \langle Q_\nu^\alpha(0) \rangle \cos \omega_\nu^\alpha t + \langle
P_\nu^\alpha(0) \rangle \frac{\sin \omega_\nu^\alpha t}{\omega_\nu^\alpha} \Big)
\end{equation}
vanish because $\langle Q_\nu^\alpha(0) \rangle = \langle P_\nu^\alpha(0)
\rangle = 0$ for thermal bath states $\rho_B^\alpha(0) \propto
\rme^{-\beta_\alpha H_B^\alpha}$, i.\,e.~the random noise forces are not biased.
In the considered case with nonthermal bath preparations the expectation values
$\langle \eta_\alpha(t) \rangle$, in general, are finite.
This leads to a finite shift of the central oscillator.
In the thermodynamic limit, where
\begin{equation}
 \langle \eta_\alpha(t) \rangle = \int_0^\infty \rmd\omega \: D_\alpha(\omega)
\lambda_\alpha(\omega) \Big( X_{\alpha,Q}(\omega) \cos \omega t +
X_{\alpha,P}(\omega) \frac{\sin \omega t}{\omega} \Big) \;,
\end{equation}
we can use the Riemann-Lebesgue lemma~\eqref{RLlemma} to find $\lim_{t \to
\infty} \langle \eta_\alpha(t) \rangle = 0$.
In conclusion, for transient times $t < \infty$ the generalization of the
initial preparation from thermal to general nonthermal states introduces a shift of the central oscillator which vanishes in the long-time limit.

The correlations of the noise forces are given by
\begin{equation}\begin{split}\label{corrfunc}
 S_{\eta_\alpha \eta_\beta}(t, s) &= \frac{1}{2} \langle \eta_\alpha(t)
\eta_\beta(s) + \eta_\beta(s) \eta_\alpha(t) \rangle - \langle \eta_\alpha(t)
\rangle \langle \eta_\beta(s) \rangle \\
 &= \sum_{\nu,\mu=1}^{N_\alpha}
\lambda_\nu^\alpha \lambda_\mu^\alpha \Big( \cos
(\omega_\nu^\alpha t) \cos (\omega_\mu^\alpha s) \sigma_{Q_\nu^\alpha
Q_\mu^\alpha} + \big[ \cos (\omega_\nu^\alpha t) \sin (\omega_\mu^\alpha s)
 \\
 &\quad + \cos (\omega_\nu^\alpha s) \sin (\omega_\mu^\alpha t) \big]
\frac{\sigma_{Q_\nu^\alpha P_\mu^\alpha}}{\omega_\mu^\alpha} + \sin
(\omega_\nu^\alpha t) \sin (\omega_\mu^\alpha s) \frac{\sigma_{P_\nu^\alpha
P_\mu^\alpha}}{\omega_\nu^\alpha \omega_\mu^\alpha} \Big) \delta_{\alpha,
\beta} \;.
\end{split}\end{equation}
Of course, $S_{\eta_\alpha \eta_\beta}(t, s) = 0$ for $\alpha \neq \beta$ due
to our assumption of factorizing states.
Starting again with the thermal bath preparation, where $(\omega_\nu^\alpha)^2
\sigma_{Q_\nu^\alpha Q_\mu^\alpha} = \sigma_{P_\nu^\alpha P_\mu^\alpha} =
\mathcal{E}_{\text{th}}(\omega_\nu^\alpha, T_\alpha) \delta_{\nu,\mu}$ and
$\sigma_{Q_\nu^\alpha P_\mu^\alpha} = 0$ we obtain
\begin{equation}\label{S_eta_therm}
 S_{\eta_\alpha \eta_\beta}^{\text{th}}(t, s) =S_{\eta_\alpha
\eta_\beta}^{\text{th}}(t-s,0) = \sum_{\nu}^{N_\alpha} \Big(
\frac{\lambda_\nu^\alpha}{\omega_\nu^\alpha} \Big)^2 \cos \omega_\nu^\alpha (t -
s) \mathcal{E}_{\text{th}}(\omega_\nu^\alpha, T_\alpha) \delta_{\alpha,\beta}
\;.
\end{equation}
This expression depends on $t-s$ only, i.\,e.~it is time-homogeneous.
In the thermal case, the fluctuating forces constitute a stationary Gaussian
process.
For a nonthermal bath preparation, the correlation functions $S_{\eta_\alpha
\eta_\beta}(t, s)$, in general, are not time-homogeneous.
Performing the thermodynamic limit, the correlation function in
Eq.~\eqref{corrfunc} assumes the form
\begin{equation}\begin{split}
 S_{\eta_\alpha \eta_\beta}(t, s) &= \int_0^\infty \rmd\omega \:
\frac{\gamma_\alpha(\omega)}{\omega} \Big( \omega^2 \cos (\omega t) \cos (\omega
s) \sigma_{\alpha,QQ}^{(1)}(\omega) + \omega \sin \omega (t + s)
\sigma_{\alpha,QP}^{(1)}(\omega) \\
 &\quad + \sin (\omega t) \sin (\omega s) \sigma_{\alpha,PP}^{(1)}(\omega)
\Big) \delta_{\alpha, \beta} \\
 &\quad + \int_0^\infty \rmd\omega_1 \int_0^\infty \rmd\omega_2 \:
D_\alpha(\omega_1) D_\alpha(\omega_2) \lambda_\alpha(\omega_1)
\lambda_\alpha(\omega_2) \\
 &\quad\times \Big( \cos (\omega_1 t) \cos (\omega_2 s)
\sigma_{\alpha,QQ}^{(2)}(\omega_1, \omega_2) + \big[ \cos (\omega_1 t) \sin
(\omega_2 s) \\
 &\quad + \cos (\omega_1 s) \sin (\omega_2 t) \big]
\frac{\sigma_{\alpha,QP}^{(2)}(\omega_1, \omega_2)}{\omega_2} + \sin (\omega_1
t) \sin (\omega_2 s) \frac{\sigma_{\alpha,PP}^{(2)}(\omega_1,
\omega_2)}{\omega_1 \omega_2} \Big) \delta_{\alpha, \beta} \;.
\end{split}\end{equation}
In the limit $t \to \infty$ and/or $s \to \infty$, the non-diagonal parts
with double frequency integrals as well as the term
involving $\sigma_{\alpha,QP}^{(1)}(\omega)$ vanish.
The remaining terms
\begin{equation}
 \int_0^\infty \rmd\omega \: \frac{\gamma_\alpha(\omega)}{\omega} \Big( \omega^2
\cos (\omega t) \cos (\omega s) \sigma_{\alpha,QQ}^{(1)}(\omega) + \sin (\omega
t) \sin (\omega s) \sigma_{\alpha,PP}^{(1)}(\omega) \Big) \delta_{\alpha, \beta}
\end{equation}
disappear as well if only one of the two variables $t$ or $s$ independently 
approaches infinity.
A finite contribution to $S_{\eta_\alpha \eta_\beta}(t, s)$ is obtained when $t$
and $s$ simultaneously approach infinity, such that
\begin{equation}
 \lim_{t \to \infty} S_{\eta_\alpha \eta_\beta}(t, t + s) = \int_0^\infty 
\rmd\omega \: \frac{\gamma_\alpha(\omega)}{\omega} \mathcal{E}_\alpha(\omega)
\cos (\omega s) \delta_{\alpha, \beta} \;.
\end{equation}
We note that this result is identical to the thermal result in
Eq.~\eqref{S_eta_therm}, provided that $\mathcal{E}_\alpha(\omega)$ is replaced
by the corresponding expression for the thermal state of a single bath.

In contrast to $\eta_\alpha(t)$, the statistical properties of the fluctuating
forces $\xi_\alpha(t)$ including the initial slip term explicitly depend on the
initial preparation of the total system including the central oscillator.
For the random forces $\xi_\alpha(t)$ to be not biased and to have
time-homogeneous correlations, the total system should be prepared in the state
$\rho(0) \propto \rho_S(0) \prod_{\alpha=1}^{N_B} \rme^{-\beta_\alpha
(H_B^\alpha + H_{SB}^\alpha)}$, where the bath contains shifted oscillators, see
Ref.~\cite{Hanggi2005} for a detailed discussion.
In the nonthermal situation, we have
\begin{equation}
 \langle \xi_\alpha(t) \rangle = \langle \eta_\alpha(t) \rangle + \langle Q(0)
\rangle \sum_{\nu=1}^{N_\alpha} \Big(
\frac{\lambda_\nu^\alpha}{\omega_\nu^\alpha} \Big)^2 \cos \omega_\nu^\alpha t
\;,
\end{equation}
and
\begin{equation}
 S_{\xi_\alpha, \xi_\beta}(t, s) = S_{\eta_\alpha, \eta_\beta}(t, s) +
\Sigma_{QQ}(0) \sum_{\nu=1}^{N_\alpha} \sum_{\mu=1}^{N_\beta} \Big(
\frac{\lambda_\nu^\alpha \lambda_\mu^\beta}{\omega_\nu^\alpha \omega_\mu^\beta}
\Big)^2 \cos (\omega_\nu^\alpha t) \cos (\omega_\nu^\beta s) \;.
\end{equation}
Performing the thermodynamic limit one notes that the additional contributions
from the initial slip term vanish in the long-time limit $t \to \infty$ and/or
$s \to \infty$.
We remark that the stationarity of the correlation functions in the long-time
limit is a consequence of the thermodynamic limit and does not rely on
ergodicity.

\end{document}